\documentclass[superscriptaddress,twocolumn,prx,showpacs,preprintnumbers,amsmath,amssymb]{revtex4}
\usepackage{graphicx,afterpage}
\usepackage{xcolor}
\usepackage[colorlinks=true,plainpages=false,linkcolor=blue,urlcolor=blue,citecolor=blue,pdfpagemode=UseNone,pdfstartview=FitBH]
{hyperref}

\begin{document}

\title{Robust Charge-Density Wave Correlations in Optimally-Doped YBa$_2$Cu$_3$O$_{\rm{y}}$}

\author{Rui Zhou}
\email{rzhou@iphy.ac.cn}
\affiliation{CNRS, LNCMI, Univ. Grenoble Alpes, INSA-T, UPS, EMFL, Grenoble, France}
\author{Igor Vinograd}
\affiliation{CNRS, LNCMI, Univ. Grenoble Alpes, INSA-T, UPS, EMFL, Grenoble, France}
\author{Hadrien Mayaffre}
\affiliation{CNRS, LNCMI, Univ. Grenoble Alpes, INSA-T, UPS, EMFL, Grenoble, France}
\author{Juan Porras}
\affiliation{Max-Planck-Institut f\"{u}r Festk\"{o}rperforschung, Heisenbergstr. 1, 70569 Stuttgart, Germany}
\author{Hun-Ho Kim}
\affiliation{Max-Planck-Institut f\"{u}r Festk\"{o}rperforschung, Heisenbergstr. 1, 70569 Stuttgart, Germany}
\author{Toshinao~Loew}
\affiliation{Max-Planck-Institut f\"{u}r Festk\"{o}rperforschung, Heisenbergstr. 1, 70569 Stuttgart, Germany}
\author{Yiran~Liu}
\affiliation{Max-Planck-Institut f\"{u}r Festk\"{o}rperforschung, Heisenbergstr. 1, 70569 Stuttgart, Germany}
\author{Matthieu Le Tacon}
\affiliation{Max-Planck-Institut f\"{u}r Festk\"{o}rperforschung, Heisenbergstr. 1, 70569 Stuttgart, Germany}
\affiliation{Institute of Solid State Physics (IFP), Karlsruhe Institute of Technology, D-76021 Karlsruhe, Germany}
\author{Bernhard Keimer}
\affiliation{Max-Planck-Institut f\"{u}r Festk\"{o}rperforschung, Heisenbergstr. 1, 70569 Stuttgart, Germany}
\author{Marc-Henri Julien}
\email{marc-henri.julien@lncmi.cnrs.fr}
\affiliation{CNRS, LNCMI, Univ. Grenoble Alpes, INSA-T, UPS, EMFL, Grenoble, France}
\date{\today}


\begin{abstract}

Charge-density wave (CDW) order is a key property of high-$T_c$ cuprates, but its boundaries in the phase diagram and potential connections to other phases remain controversial. We report nuclear magnetic resonance (NMR) measurements in the prototypical cuprate YBa$_2$Cu$_3$O$_{\rm y}$ demonstrating that short-range static CDW order remains robust at optimal doping ($p=0.165$), exhibiting a strength and temperature dependence in the normal state similar to those observed at $p\simeq0.11$ \textcolor{black}{in the underdoped regime}. For \textcolor{black}{an overdoped sample with} $p=0.184$, we detect no static CDW down to $T=T_c$, though weak CDW order plausibly emerges below $T_c$. More broadly, we argue that both quenched disorder and competition with superconductivity influence the apparent boundary of the CDW phase, likely causing an underestimation of its intrinsic extent in doping. These findings challenge the view that the CDW phase boundary lies below $p^*\simeq0.19$, widely regarded as the critical doping where the pseudogap phase ends in YBa$_2$Cu$_3$O$_{\rm{y}}$.
\end{abstract}

\maketitle

{\bf Introduction}~--~The most significant advance in research on cuprate superconductivity over the last decade is arguably the realization that charge density wave (CDW) order is not only pervasive across all cuprate superconductors~\cite{Comin2016,Frano2020,Uchida2021,Hayden2023} but also the ground state of \textcolor{black}{the simplest version of the Hubbard model on a square lattice within a parameter regime relevant to real, superconducting compounds (namely a hole density $p\sim 0.1$ per unit cell and a ratio $U/t \simeq 8$ between of the on-site Coulomb repulsion~$U$ to the nearest-neighbor hopping~$t$)~\cite{Zheng2017}.  }

\textcolor{black}{Nonetheless, the CDW has not yet revealed all its secrets. One of remaining outstanding issues is whether the boundaries of the CDW phase correlate with other key phenomena in the phase diagram as a function of $p$ and temperature $T$, including superconductivity itself. More specifically, a key question is whether the critical doping up to which the CDW exists on the high-doping side coincides with any of the following: the optimal doping for superconductivity at $p \simeq 0.16$, the end of the superconducting dome at $p \simeq 0.3$, the end of the pseudogap phase at $p^* \simeq 0.19$ (and concomitant onset of the strange metal phase), or none at all. Experiments across different cuprate families and using different probes have so far provided conflicting answers~\cite{Fujita2014,Peng2018,Miao2021,Li2021,Li2023,Song2023,Lu2022,vonArx2023}.}

\textcolor{black}{In fact, the inherent difficulty of determining CDW phase boundaries is compounded by two key factors in the cuprates: quenched disorder and competition of orders. Dopant-induced disorder is ubiquitous and a short CDW correlation length (typically $\xi_{\rm CDW} < 20$ lattice units) is the norm rather than the exception. Taking disorder into account has proved crucial to understand the particularly intricate CDW phenomenology~\cite{Kivelson2003,Nie2014,LeTacon2014,Wu2015,Mesaros2016,Caplan2017,Mukhopadhyay2019,Lee2022}. In addition, disorder contributes to structural and electronic inhomogeneity that frequently complicates experimental detection of a density wave. Last, the competing effect of superconductivity introduces another difficulty at low temperatures, as it also degrades both the amplitude and the spatial coherence of the CDW, making it even more challenging to detect experimentally~\cite{Wu2011,Ghiringhelli2012,Chang2012,Wu2013}. }

\textcolor{black}{Addressing these issues in YBa$_2$Cu$_3$O$_{\rm y}$ (YBCO) is particularly important for two reasons. First, YBCO hosts what is arguably the prototypical CDW in the normal state of underdoped cuprates ($p \simeq 0.11$--$0.13$): its short-range order is not intertwined with any spin order and exhibits the longest correlation length among cuprates -- extending up to approximately 17 lattice spacings, or 5 to 6 times the CDW period~\cite{Ghiringhelli2012,Chang2012,Blanco-Canosa2014,Huecker2014}. Second, the extent of the short-range CDW phase is central to interpreting Hall effect measurements, which reveal a striking change in carrier density from approximately $p$ below optimal doping ($p=0.16$) to $1+p$ above the pseudogap boundary $p^*=0.19$~\cite{Badoux2016,Proust2019,Zou2017,Morice2017,Caprara2017,Sharma2018}. An influential view has been that CDW order terminates at $p \simeq 0.16$, that is clearly before the pseudogap boundary $p^*$, and that material-specific density-wave orders play no role in the $p$ to $1+p$ transition~\cite{Badoux2016,Proust2019}. This conclusion was based on two X-ray studies of near-optimally doped YBCO that found CDW peaks to be weak~\cite{Blanco-Canosa2014} or even absent~\cite{Huecker2014}. }

\textcolor{black}{Underdoped YBCO also exhibits long-range CDW order in intense magnetic fields~\cite{Wu2011,Wu2013,Gerber2015,Julien2015,Vinograd2021} or under uniaxial strain~\cite{Kim2018}. However, this phase is specific to YBCO, has less evident impact on normal-state properties than its short-range counterpart~\cite{Zhou2017b,Cyr2017,Laliberte2018,Zhou2024}, and -- near optimal doping, where the superconducting upper critical field exceeds 100~T -- is inaccessible to nuclear magnetic resonance (NMR), if it even exists~\cite{Nakata2025}. We therefore investigated the short-range CDW in YBCO using NMR, and} show that static charge modulations are clearly present at optimal doping and are unlikely to vanish before $p^*$.


{\bf NMR spectra}~--~We studied single crystals of YBa$_2$Cu$_3$O$_{6.92}$ (hole doping $p=0.165\pm0.005$) and YBa$_2$Cu$_3$O$_{6.98}$ ($p=0.184\pm0.005$) -- see End Matter section for details on the samples. Their quality is attested by the sharp NMR lines, though oxygen disorder shows up through lineshape asymmetry and additional resonances (see End Matter), as expected from the presence of vacancies with respect to the full chains of YBa$_2$Cu$_3$O$_{7.0}$. In the remainder of this Letter, we focus on results from O(3) lines, corresponding to planar oxygen atoms involved in bonds oriented along the $b$ axis. Indeed, the complete separation between the standard O(3) and vacancy-induced O(3v) lines allows for the most precise measurements. O(2) lines yield consistent results, though with larger error bars.


   \begin{figure}[t!]
 \includegraphics[width=8.4cm]{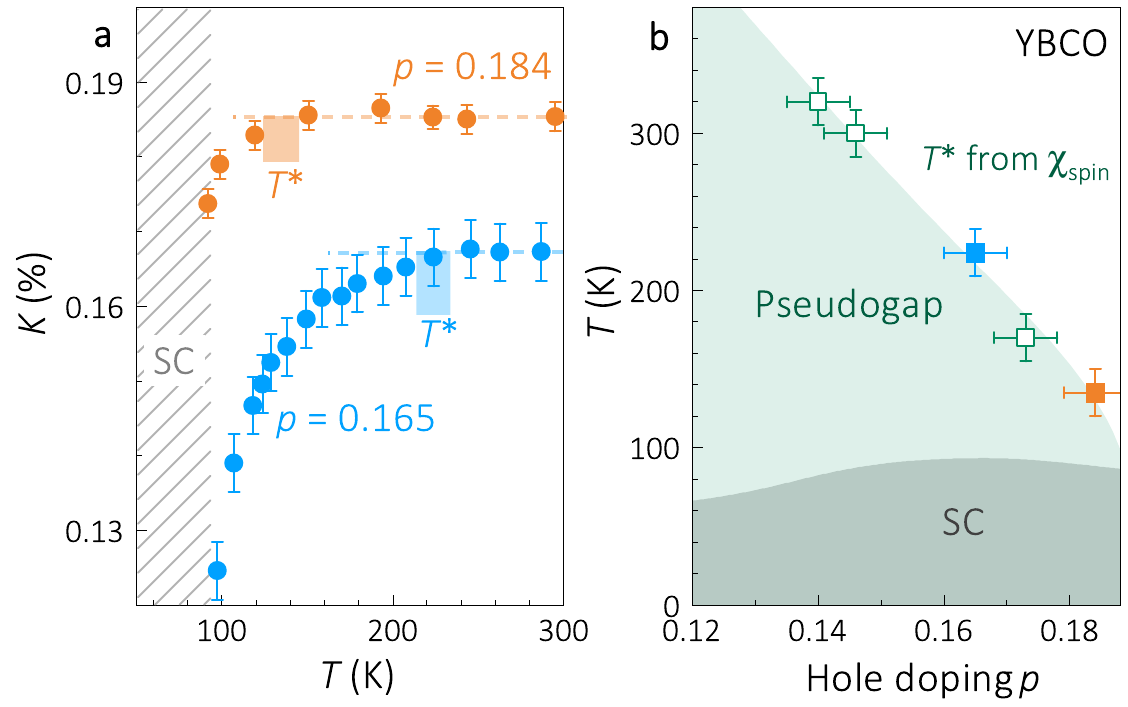}
  \caption{\label{shift} Pseudogap temperature $T^*$. (a) Knight shift $K \propto \chi_{\rm spin}$ vs. temperature for O(3) sites in $p=0.165$ (blue) and $p=0.184$ (orange) samples. The rectangles mark the onset of visible decrease of $K$ upon cooling, which defines the pseudogap onset temperature $T^*$. For $p=0.165$, $K$ shows a sharper decrease below 130 - 150~K, i.e. well below $T^*$, which is also where CDW correlations become significant (Fig.~\ref{width}). This correlation is consistent with recent work indicating that short-range CDW order contributes to the reduction in $\chi_{\rm spin}(T)$~\cite{Zhou2024}. (b) Phase diagram showing that the $T^*$ values are consistent with literature data at other doping levels in YBCO~\cite{Alloul2012}. Notice that superconducting fluctuations are expected to be negligible above approximately 100~K for near-optimally-doped YBCO~\cite{Grbic2011}.}
\end{figure} 

{\bf Pseudogap temperature~--~}From the spectra, we extract the Knight shift $K$, which is proportional to the local uniform ($q=0$) static ($\omega = 0$) spin susceptibility $\chi_{\rm spin}$ (the orbital contribution being negligible for $^{17}$O~\cite{Zhou2024}). As shown in Fig.~\ref{shift}a, $K$ is constant at high temperatures (at least within the error bars, on this $T$ range) and decreases upon cooling. Defining the pseudogap onset temperature $T^*$ as the temperature below which $K$ decreases~\cite{Alloul1989}, the $T^*$ values determined in this way agree with previous NMR studies~\cite{Alloul2012} and parallel the doping dependence of the pseudogap temperature scale in specific heat measurements~\cite{Loram2001}. Both our samples are in the pseudogap regime but it is important to note that $T^* = 130 \pm 15$~K for 0.18 is significantly lower than for $p=0.165$ ($T^* = 230 \pm 15$~K).

It is important to note that the shielding of radiofrequency prevents NMR measurements from being performed deep in the superconducting state of these two samples. Experiments below $T_c$ would require aligned-powder samples or steady magnetic fields beyond the current state of the art.

\begin{figure}[t!]
 \includegraphics[width=7cm]{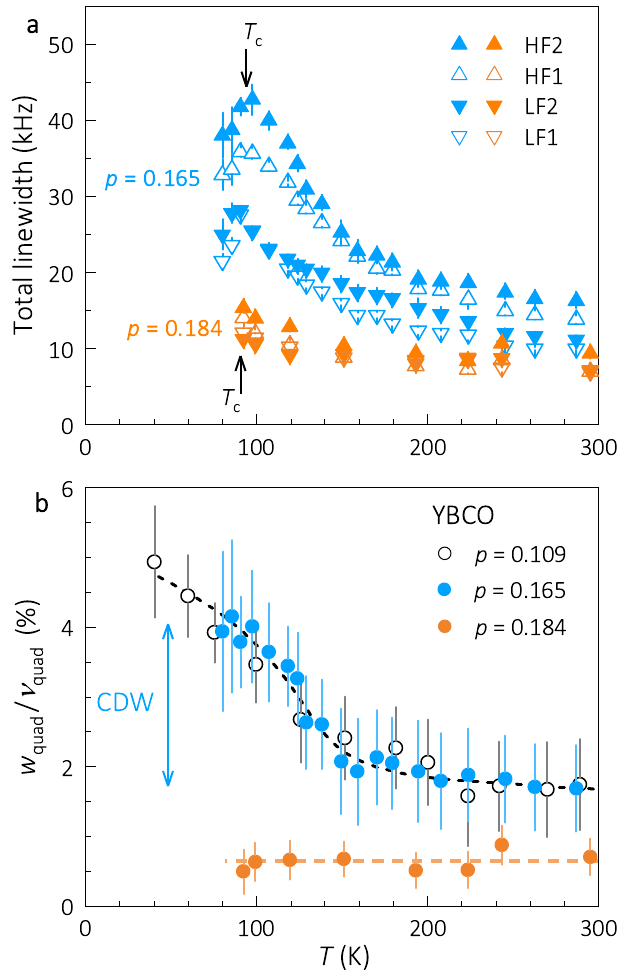}
  \caption{\label{width} Evidence for static CDW. (a) Full width at half maximum of each of the four satellite lines, for $p=0.165$ (blue) and $p=0.184$ (orange) samples. (b) Quadrupole contribution to the width $w_{\rm quad}$ relative to the quadrupole frequency $\nu_{\rm quad}=(\nu_{\rm HF1}-\nu_{\rm LF1})/2$ where $\nu_{\rm HF1}$ ($\nu_{\rm LF1}$) is the resonance frequency of the first high (low) frequency satellite. The $T$ dependence of $w_{\rm quad}$ directly reflects the growth of CDW correlations. See End Matter for details on the analysis. The dashed lines guide the eye.}
\end{figure}

{\bf Short-range CDW order~--~}Fig.~\ref{width}a shows the temperature ($T$) dependence of the width of each of the four satellite lines of O(3). At 300~K, the lines are already broader for $p = 0.165$ than for $p = 0.184$, which is consistent with the greater O disorder in the former. Upon cooling, the lines broaden significantly more for $p = 0.165$, and this broadening is more pronounced for high-frequency (HF) satellites than for low-frequency (LF) ones (compare HF2 with LF2 and HF1 with LF1 in Fig.~\ref{width}a). As explained in our earlier works on the short-range CDW in the normal state of underdoped YBCO~\cite{Wu2015,Vinograd2019}, this behavior indicates that the $T$ dependent broadening arises from a joint spatial heterogeneity in the Knight shift $K$ (thus in the uniform magnetic susceptibility) and in the quadrupole parameters (thus in the electric field gradient - EFG). Such coupled effects are expected in presence of a periodic modulation of the charge density, and are indeed observed in underdoped YBCO~\cite{Wu2015}. In principle, spatial inhomogeneity of the carrier concentration, on scales of nanometers or larger, could produce similarly-coupled distributions. However,  it would not lead to the marked temperature dependence of the broadening we observe above $T_c$. Therefore, independently of any model, the $T$ dependent broadening affecting high-frequency satellites more strongly than low-frequency ones provides qualitative evidence for static CDW modulations at $p = 0.165$ in YBCO.


For quantitative insight, it is essential to separate the two channels of line broadening, namely, the 'magnetic' component $w_{\rm magn}$ and the 'quadrupole' component $w_{\rm quad}$. $w_{\rm magn}$ represents the width of the distribution of Knight shifts or, equivalently, of local magnetic susceptibilities. $w_{\rm quad}$ corresponds to the distribution of quadrupole frequencies $\nu_{\rm quad}=(\nu_{\rm HF1}-\nu_{\rm LF1})/2$ where $\nu_{\rm HF1}$ ($\nu_{\rm LF1}$) is the resonance frequency of the first high (low) frequency satellite. $\nu_{\rm quad}$ is itself a function of the diagonal elements of the EFG tensor and of the field orientation. Separating  $w_{\rm magn}$ and $w_{\rm quad}$ is important because it is only for $w_{\rm quad}$ that the $T$-dependent part of broadening is directly related to the CDW. $w_{\rm magn}(T)$, on the other hand, is contributed by both the CDW and the magnetic response to defects. This latter is a distribution of Knight shifts that widens upon cooling due to staggered-magnetization puddles forming around non-magnetic defects~\cite{Julien2000,Ouazi2006,Alloul2009,Chen2009}. 

As detailed in End Matter section, we infer $w_{\rm quad}$ and $w_{\rm magn}$ from the same model used in underdoped YBCO~\cite{Wu2015}, where it was shown that the $T$ dependence of $w_{\rm quad}$ matches the intensity of CDW peaks in X-ray scattering (although, in principle, both the amplitude and the correlation length of the CDW contribute to $w_{\rm quad}$). For the purpose of comparing different doping levels, we consider the dimensionless ratio $w_{\rm quad}/\nu_{\rm quad}$, i.e. the width of the distribution relative to its central value $\nu_{\rm quad}$ that depends on doping~\cite{SI}. The extracted~broadening $w_{\rm quad}/\nu_{\rm quad} (T)$, shown in Fig.~\ref{width}b, is our main result, revealing three main features:

1) For $p=0.165$, local CDW order becomes evident below approximately 150~K, though experimental uncertainty prevents us from determining whether it emerges sharply at $\sim$160~K or rises gradually from higher temperatures. We speculate that the data actually follow an S-shaped dependence with an inflection point around 130~K, which makes it impossible to unambiguously define an onset temperature $T_{\rm CDW}$.

2) The broadening for $p=0.165$ has the same amplitude and $T$ dependence as in ortho-II YBCO with $p = 0.11$, even though larger values of $w_{\rm quad}/\nu_{\rm quad}$ are ultimately reached below 90~K for the latter as its lower $T_c$ of 60~K provides a wider temperature window in the normal state. Since error bars are relatively large \textcolor{black}{and the model used for the analysis is inevitably simplified}, we refrain from claiming that CDW order is strictly identical for both $p=0.11$ and $0.165$. However, it is certainly quite comparable.

3) For $p = 0.184$, no $T$-dependent quadrupole broadening is observed \textcolor{black}{within error bars}, meaning that local CDW order is absent down to at least 90~K.


{\bf NMR vs. X-rays~--~}Our finding of robust short-range CDW order below about 150~K at $p=0.165$ seemingly conflicts with X-ray scattering finding weak CDW peak intensity below 100~K at $p=0.163$~\cite{Blanco-Canosa2014} or even no CDW at $p=0.165$~\cite{Huecker2014}. One possible explanation is that a local probe like NMR is more sensitive than a scattering technique if the CDW shows strong spatial inhomogeneity. However, there is no evidence that this is relevant here. In fact, we shall now argue that a critical examination of X-ray scattering data does not suggest any serious discrepancy with our NMR data. 

We first remark that the absence of CDW at $p=0.165$ for H\"{u}cker et al.~\cite{Huecker2014} is difficult to reconcile with the results of Blanco-Canosa et al.~\cite{Blanco-Canosa2014} as this would entail an abrupt collapse of the CDW above $p \simeq 0.163$ that appears unphysical. The absence of detectable CDW in ref.~\cite{Huecker2014} may instead find a natural explanation in two factors: the lower sensitivity of hard X-ray diffraction compared to the resonant soft X-ray scattering used in ref.~\cite{Blanco-Canosa2014} and an actual doping slightly exceeding the estimated $p=0.165$ so that the temperature $T_{\rm CDW}$ at which CDW begins to be detected in X-ray scattering is actually lower than $T_c$. In this case, CDW would be weakened by superconductivity and thus even more difficult to detect. Indeed, the literature data~\cite{SI} suggest that $T_{\rm CDW}$ extrapolates to below $T_c$ just above $p=0.16$ (see also~\cite{SI}). Small doping differences around this special point may then lead to quite different results.
 
Next, we emphasize that the strength of CDW correlations should be compared across doping levels at the same temperature, rather than at their respective $T_c$, since superconductivity affects the strength of the CDW. The difference is notable: for underdoped YBCO with $T_c=60$~K, the peak intensity in X-ray scattering is about 70\% smaller at 90~K than at $T_c\simeq$~60~K~\cite{Ghiringhelli2012,Chang2012,Huecker2014,Blanco-Canosa2014,Comin2015,Wandel2022}. Furthermore, a rigorous comparison should be based not just on intensities at the CDW wave vector but on integrated intensities (peak intensity multiplied by the width in three dimensions), a quantity proportional to the square of the modulation amplitude~\cite{Huecker2014}. In YBCO, the loss of peak intensity from $p=0.11$ to $p=0.16$~\cite{Blanco-Canosa2014} is, at least partially, compensated by the broadening as the in-plane correlation length shortens by a factor about 2.5 between $p=0.12$ and 0.16. Such quantitative comparison of CDW strength between underdoped and optimally doped YBCO was actually performed for Nd$_{1+x}$Ba$_{2-x}$Cu$_3$O$_{7-\rm{\delta}}$ (NBCO) thin films~\cite{Arpaia2019}. Although this aspect was not highlighted in the original analysis, the data (reproduced here in~\cite{SI}) further support the notion of robust CDW correlations at optimal doping: the elastic peaks for $p=0.17$ are broader than for $p=0.11$ (i.e. the CDW is shorter ranged for $p=0.17$), the integrated areas are similar for both dopings down to 90~K, the $T_c$ value for $p=0.17$. However, given potential subtle differences in doping, disorder, and structure between NBCO films and YBCO crystals, refined X-ray investigations in near optimally doped YBCO remain desirable.

{\bf Where is the CDW endpoint?~--~}The robustness of the CDW at $p = 0.165$ does not indicate at which doping level it disappears. While signatures of a static CDW are absent for $p = 0.184$ (normal-state NMR data in Fig.~\ref{width}b), \textcolor{black}{0.185~\cite{Arpaia2023}}  and $p = 0.189$~\cite{Blanco-Canosa2014}), it remains possible that CDW correlations were simply too weak or emerged at too low temperatures to be detected. Two observations support this possibility. First, a fit of $T_{\rm CDW}$ data from refs.~\cite{Blanco-Canosa2014,Huecker2014} to a parabolic $p$ dependence suggests that $T_{\rm CDW}$ drops steeply above $p=0.16$ and vanishes at $p=0.19 - 0.20$~\cite{SI}. This should be considered as a lower bound since, as we argue later,  $T_{\rm CDW}$ values near $p=0.16$ are likely underestimated.  Second, a CDW transition at 35~K --well below $T_c$-- was reported in YBa$_2$Cu$_3$O$_7$ ($p\simeq0.183$)~\cite{Kramer1999}. While later work suggested this effect could arise from charge modulations in the chains influencing the CuO$_2$ planes~\cite{Grevin2000}, the modern understanding that the planes exhibit an intrinsic CDW susceptibility invites a reconsideration of these results. 

At any rate, given the proximity of $p\simeq 0.16$ to the pseudogap boundary $p^* \simeq 0.19$~\cite{Loram2001}, it is highly unlikely that the CDW phase terminates significantly before $p^*$. In this regard, we note that, according to ref.~\cite{Zou2017}, the magnetotransport properties of YBCO are not inconsistent the presence of CDW correlations at $0.16 \leq p \leq 0.19$. \textcolor{black}{We also emphasize that our discussion focuses on static CDW correlations as a proxy for the presence of CDW order at $T=0$, disregarding charge-density fluctuations~\cite{Arpaia2019}. These persist at higher doping levels, have minimum energy at $p^*$ and potentially play a significant role in various aspects of cuprate physics~\cite{Arpaia2023}.}


   \begin{figure}[t!]
 \includegraphics[width=6.5cm]{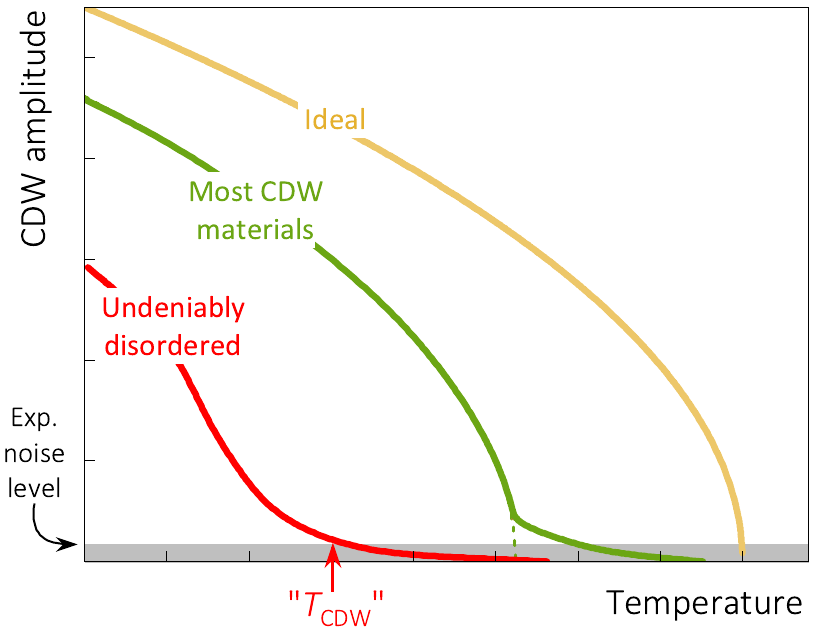}
  \caption{\label{disorder} Illustration of how disorder affects the $T$ dependence of the CDW amplitude (inspired by refs.~\cite{Straquadine2019,Mallayya2024} on an unidirectional CDW in a layered system) and how this impacts the determination of an apparent onset temperature $T_{\rm CDW}$, which can strongly depend on the signal-to-noise ratio (i.e. experimental sensitivity). Many CDW materials, even among those nominally disorder-free, show a small pretransitional tail at high temperature due to disorder (green curve). Cuprates and other materials with off-stoichiometric dopants more often follow the red curve (if not for competing effects due to superconductivity at $T \ll T_{\rm CDW}$).}
\end{figure} 

{\bf Concluding remarks~--~}Whether the CDW terminates at or above the pseudogap end point $p^*$ has been discussed in multiple cuprate works, with arguments in one direction~\cite{Fujita2014,Arpaia2023,Ramshaw2015} or the other~\cite{Peng2018,Miao2021,Tam2022,Li2023}. Our aim here is not to advocate for one viewpoint over the other but to underscore that the question of CDW phase boundaries, whether as a function of $p$ or $T$, is more intricate than often assumed. We therefore conclude this Letter with a set of considerations about the effect of disorder, intended to place the debate on a more solid footing --while recognizing that disorder is a more complex issue than discussed here and that a number of other complications are important, among which is the presence of superconductivity that potentially masks the intrinsic $T=0$ boundaries of competing phases, as discussed above and recently highlighted in La$_{\rm 2-x}$Sr$_{\rm x}$CuO$_4$~\cite{Frachet2020,Vinograd2022}. 

\textcolor{black}{Let us first recall very generally that the effect of quenched disorder on CDWs has been theoretically anticipated~\cite{McMillan1975,Imry1975,Sham1976,Efetov1977,Fukuyama1978} and experimentally documented since the 70s~(see refs. \cite{Wilson1974,Berthier1978,Skripov1983,Weitering1999,Ghoshray2009,Chatterjee2015,Liu2021,Straquadine2019,Yue2020,Mallayya2024} for layered systems though numerous examples also exist in 1D systems~\cite{Pouget2016}). Disorder can reduce the transition temperature, suppress the amplitude, and shorten the correlation length of the CDW. It also nucleates local static modulations of the charge density -- an effect sometimes described as the 'pinning of CDW fluctuations'. While the magnitude of these effects may depend on microscopic details of the CDW system and of the nature of disorder, the distinction between short-range and long-range order can ultimately become blurred. As a result, thermodynamic signatures of the transition may be smeared out as the temperature dependence of the order parameter becomes more gradual, as illustrated in Fig.~\ref{disorder}. }

Because the growth of the CDW order parameter upon cooling in the normal state is gradual, there is an extended temperature range where experimental signatures of the CDW remain weak across all probes. As a result, the determination of an onset temperature is inherently biased by the signal-to-noise ratio ($S/N$) of the measurements (see Fig.~\ref{disorder}). In fact, the very concept of an onset temperature $T_{\rm CDW}$ becomes questionable and a Curie-Weiss-type temperature dependence provides a possible description of the data in the purported onset region (see ref.~\cite{Vinograd2019} for an example in YBCO). 

These considerations have direct implications for the cuprates, including YBCO. First, $T_{\rm CDW}$, determined from the peak intensity in X-rays, may be underestimated when moving away from $p=0.12$ due to poorer $S/N$. This possibly causes the dome-shaped $T_{\rm CDW}$ and, consequently, the apparent discrepancy between this dome and the present NMR findings, showing identical $T$ dependence for $p=0.11$ and $p=0.165$. In fact, $T_{\rm CDW}$ in La$_{\rm 2-x}$Sr$_{\rm x}$CuO$_4$ clearly depends on experimental sensitivity (i.e., $S/N$)~\cite{Wang2020}, a possibility already suggested in refs.~\cite{Wu2015,Mukhopadhyay2019}. \textcolor{black}{Also consistent with this, the latest generation of resonant inelastic X-ray scattering measurements report larger $T_{\rm CDW}$ values in YBCO~\cite{Arpaia2019,Wahlberg2021,Wandel2022}.}

In theory, the gradual temperature dependence of the CDW amplitude may arise from the competition with superconducting fluctuations~\cite{Hayward2014,Achkar2014,Caplan2017,Orgad2025}. However, we find that disorder offers a more likely explanation, because (i) such $T$ dependence is widespread in CDWs~\cite{Berthier1978,Skripov1983,Weitering1999,Ghoshray2009,Kogar2017,Straquadine2019,Feng2023,Mallayya2024}, (ii) dopant-induced disorder is present in virtually all cuprates, and (iii) a dependence of $T_{\rm CDW}$ on disorder was recently found to explain data in La$_{\rm 1.8-x}$Eu$_{0.2}$Sr$_{\rm x}$CuO$_4$~\cite{Lee2022} and YBa$_2$Cu$_4$O$_8$~\cite{Betto2025}. In both cases, it was found that, in the disorder-free limit, $T_{\rm CDW}$ is on the order of $T^*$ \textcolor{black}{(see also ref.~\cite{Wahlberg2021})}. 

\textcolor{black}{The idea that the pseudogap and CDW are distinct phases in YBCO has been largely shaped by two key observations: (1) $T_{\rm CDW}$ is significantly lower than $T^*$~\cite{Ghiringhelli2012,Chang2012,Huecker2014,Blanco-Canosa2014}, and (2) the dome-shaped $p$ dependence of $T_{\rm CDW}$~\cite{Blanco-Canosa2014} contrasts with the monotonic decrease of $T^*$ (Fig.~\ref{shift}b). The above discussion shows both points are actually much less solid than initially thought.} As there is a logical parallel between the $p$ and $T$ dependencies, this strengthens the conclusion that the CDW phase is unlikely to disappear before $p^*$ in YBCO. 

Still, it is also important to recognize that a sharply defined endpoint for CDW order as a function of hole doping may simply not exist. The coexistence of spatially separated regions with and without static CDW order -- often challenging to identify experimentally -- remains a distinct possibility beyond optimal doping.


\vspace{-0.5cm}
\section*{Acknowledgements}
We thank Riccardo Arpaia, Giacomo Coslovich, Giacomo Ghiringhelli, David LeBoeuf, Matteo Minola, Dror Orgad for valuable discussions.

Work in Grenoble was supported by the Laboratoire d'excellence LANEF in Grenoble (ANR-10-LABX-51-01). Part of this work was performed at the Laboratoire National des Champs Magn\'etiques Intenses, a member of the European Magnetic Field Laboratory (EMFL). 

\vspace{-0.5cm}

\bibliography{YBCOPTIMAL}

\clearpage

\begin{figure*}
\hspace*{-1cm}
\includegraphics[width=18cm]{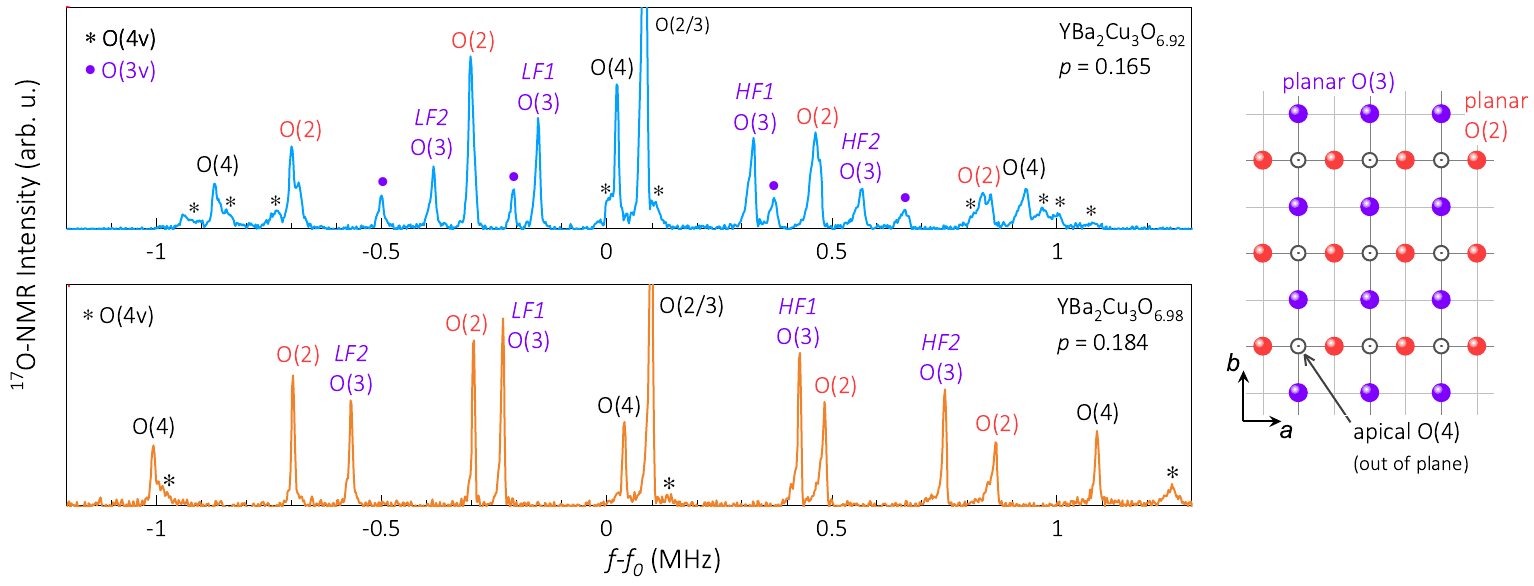}
 \caption{$^{17}$O NMR spectra for YBCO $p=0.165$ at 245~K (top) and $p=0.184$ at 151~K (bottom). Because $^{17}$O has a nuclear spin $I=5/2$, there are five resonance lines for each crystallographic site: a central line as well as two low-frequency satellites (named LF1 and LF2) and two high-frequency satellites (HF1 and HF2).  For the ideal ortho-I structure, all chains are oxygen full and there are four different crystallographic sites: the apical O(4), the chain O(1) (not seen here due to low $^{17}$O concentration \textcolor{black}{and presumably large broadening} at this site) and two planar sites, O(2) and O(3), corresponding to the bonds parallel to the $a$ and $b$ crystallographic axes, respectively, as shown in the top view of a CuO$_2$ plane (right). For $p=0.165$, the magnetic field $B\simeq 9$~T is tilted within the $bc$ plane by an angle $\theta= 16^\circ$ from the $c$ axis to maximize the separation between the regular O(3) lines and the O(3v) lines associated with vacancy nearest neighbors ($\theta= 12^\circ$ for $p=0.184$). The integrated intensity of O(3v) is $22\pm5$\% of the total O(3)+O(3v) intensity, roughly consistent with the expect 16\% of O(3) first neighbors to $\delta_{\rm v}=0.08$ isolated vacancies. O(3v) and O(4v) sites are not clearly observed in our previous works on underdoped YBCO~\cite{Wu2013,Wu2015,Zhou2017PRL,Vinograd2021,Vinograd2019}. The asymmetric shape of the lines, also due to oxygen disorder, is distinct from the $B$ and $T$ dependent asymmetry observed for the long-range ordered CDW~\cite{Zhou2017PRL}. The LF2 and HF2 satellites of O(4) are off-scale due to larger quadrupole coupling than other sites. The line assignment is based on earlier works (\cite{Wu2015} and refs. therein, and \cite{Reichardt2018} for a different perspective on the O(2)-O(3) splitting). The working frequency $f_0$ is 52.0138~MHz.}
\label{spectra}	
\end{figure*}
%

\section*{\large End Matter}


{\bf Samples~--~}High-quality single crystals of YBa$_2$Cu$_3$O$_{\rm y}$ were grown using a flux method~\cite{Blanco-Canosa2014}. They were subsequently annealed in 70\%-enriched $^{17}$O gas. The $^{17}$O substitution was confirmed by the isotope effect on O(2)-O(3) and O(4) related phonons modes measured in Raman scattering. \textcolor{black}{Notice that the concentration of $^{17}$O is smaller at the chain site as the final annealing to fine-tune the doping level is performed under low $^{17}$O partial pressure}. The studied crystals have a nominal composition y~= 6.92 and 6.98 and their hole doping $p$ is deduced from the values of the superconducting transition temperature $T_c$ (Fig.~\ref{squid}, Table~1 and ref.~\cite{Liang2006}), the $c$ axis parameter~\cite{Liang2006} and the measured quadrupole frequency $\nu_{\rm quad}$ (Fig. \ref{nuq-vs-p} for the $y=6.92$ sample), with an estimated uncertainty of $\Delta p=\pm0.005$.

\begin{table}[htbp]
\centering
\label{table1}
\begin{tabular}{ccccc}
\hline
&  \hspace{0.2cm} O content \hspace{0.2cm} & \hspace{0.2cm} Hole content  \hspace{0.2cm} & \hspace{0.2cm} Oxygen \hspace{0.2cm}  & \hspace{0.2cm} $T_c$ \hspace{0.2cm}  \\
& (nominal) & ($p$) & order & ($B$ = 0) \\
\hline\hline
& 6.92  &  $0.165\pm0.005$     &  ortho-I    &  93.0~K  \\
& 6.98  &  $0.184\pm0.005$    &  ortho-I  &  91.6~K  \\
\hline
\end{tabular}\\
\caption{Sample properties. Measurements of the the superconducting transition temperature $T_c$ are shown in Fig.~\ref{squid}.}
\end{table}

{\bf NMR experiments~--~}Standard spin-echo techniques were used with a laboratory-built heterodyne spectrometer. Spectra were obtained at fixed magnetic field by adding Fourier transforms of the spin-echo signal recorded for regularly spaced frequency values. Magnetic fields were provided by a high homogeneity 9~T superconducting magnet. Field values were calibrated using the resonance frequency of $^{27}$Al in thin Al foils ($^{27}K = 0.164\%$ at 300~K). $^{17}$O Knight shift values are given with respect to the reference frequency of the bare nucleus. 

{\bf NMR spectra and oxygen disorder}~--~In the so-called O-I structure of YBa$_2$Cu$_3$O$_{\rm y}$, the CuO$_{\rm{y}-6}$ chains are nearly oxygen-full and contain an average number $\delta_{\rm V}=7-{\rm y}$ of oxygen vacancies. These vacancies modify the electric field gradient at all nearby O and Cu sites, both in-plane and out-of-plane, resulting in distinct NMR lines for sites that are nearest neighbors to vacancies~\cite{Wu2016}.  For the $y=6.92$ sample, such lines are clearly observed (Fig.~\ref{spectra}a) as $\delta_{\rm V}=0.08$ is not small. For $y=6.98$, $\delta_{\rm V}=0.02$ and the nearest-neighbor lines are unresolved (except for the apical O) but the lineshapes are asymmetric instead (Fig.~\ref{spectra}b). 

{\bf Fitting NMR spectra~--~}To fit the asymmetric O(3) NMR lines depicted in Fig.~\ref{spectra-sup}, we employed an asymmetric Pseudo-Voigt function, as described by Schmid {\it et al.}~\cite{Schmid2014}:

\begin{equation}
\begin{split}
I(f) = m \frac{1}{2 \pi} \frac{w(f)}{\left(w(f) / 2\right)^{2}+4 (f-f_0)^{2}}+\\
(1-m) \sqrt{\frac{4 \ln (2)}{\pi w(f)^{2}}} \exp \left[-\left(4 \ln (2) /w(f)^{2}\right) (f-f_0)^{2}\right]
\label{PsdV}	
\end{split}
\end{equation}
where the effective linewidth 
\begin{equation} 
w(f)=\frac{2w_0}{1+\exp(-a(f-f_0))}
\end{equation}
induces asymmetry with respect to the standard Pseudo-Voigt function~\cite{Schmid2014} and $m$ is the relative weight of the Lorentzian part. As shown in Fig.~\ref{spectra-sup}, our fitting results exhibit high quality, enabling us to determine the full width at half maximum (FWHM) $w$ of each quadrupole line (Fig.\ref{spectra}a).


{\bf Linewidth analysis~--~}\textcolor{black}{In order to determine the quadrupole (i.e., electric field gradient, EFG) and magnetic (i.e., Knight shift) contributions to the line width, we follow the procedure of Ref.~\cite{Wu2015}. This approach was validated in underdoped YBCO, where the extracted temperature dependence of the quadrupole broadening agrees with X-ray scattering results, and the model correctly accounts for the unusual field dependence of the widths of various $^{17}$O NMR lines~\cite{Wu2015}.}

\textcolor{black}{Specifically, we assume that the different contributions to the line width are independent and combine in quadrature (i.e., are added in quadrature, which in principle is strictly valid for Gaussian broadening only), with one exception: the magnetic and quadrupole broadenings arising from spatial charge density modulations. These two are not independent, since they are spatially correlated: regions with higher (lower) local charge density exhibit both larger (smaller) Knight shift and larger (smaller) quadrupole frequency. Because the Knight shift moves all NMR lines in the same direction, while the quadrupole shift affects their relative spacing, these two effects add for the high-frequency satellites and subtract for the low-frequency satellites before being combined with the other broadening terms. This correlation explains the marked difference in width between the high- and low-frequency satellites. This is a property of the CDW in YBCO, no matter whether it is long-ranged~\cite{Wu2011,Wu2013,Vinograd2021} or short-ranged~\cite{Wu2015}.}

\textcolor{black}{The total width of the different $^{17}$O NMR satellite lines is then expressed as:
\begin{eqnarray}
{{w}_{\text{HF2}}}^2 &=& \left( 2{{w}_{\text{quad}}^0}\right)^{2}+\left({{w}_{\text{magn}}^0}\right)^{2}+{{\left( w_{\text{magn}}^{\rm cdw}+2w_{\text{quad}}^{\rm cdw} \right)}^{2}}\nonumber\\
{{w}_{\text{HF1}}}^2 &=& \left( {{w}_{\text{quad}}^0}\right)^{2}+\left({{w}_{\text{magn}}^0}\right)^{2}+{{\left( w_{\text{magn}}^{\rm cdw}+w_{\text{quad}}^{\rm cdw} \right)}^{2}}\nonumber\\
{{w}_{\text{LF1}}}^2 &=& \left({{w}_{\text{quad}}^0}\right)^{2}+ \left({{w}_{\text{magn}}^0}\right)^{2}+{{\left( w_{\text{magn}}^{\rm cdw}- w_{\text{quad}}^{\rm cdw} \right)}^{2}}\nonumber\\
{{w}_{\text{LF2}}}^2 &=& \left( 2{{w}_{\text{quad}}^0}\right)^{2}+ \left({{w}_{\text{magn}}^0}\right)^{2}+{{\left( w_{\text{magn}}^{\rm cdw}-2w_{\text{quad}}^{\rm cdw} \right)}^{2}}\nonumber
\end{eqnarray}
where one finds:}

\textcolor{black}{$\bullet$~Two independent terms: $w_{\text{quad}}^0$ and $w_{\text{magn}}^0$, representing the intrinsic quadrupolar and magnetic broadenings, respectively. Note that ${w}_{\text{magn}}^0$ is not necessarily proportional to the average Knight shift $K$. This is because, in the presence of antiferromagnetic correlations, field-induced staggered moments can form around non-magnetic defects~\cite{Julien2000,Ouazi2006,Alloul2009,Chen2009}, increasing ${w}_{\text{magn}}^0$ at low temperatures even as $K$ decreases due to the pseudogap.}

\textcolor{black}{$\bullet$~Two correlated terms: $w_{\text{quad}}^{\text{cdw}}$ and $w_{\text{magn}}^{\text{cdw}}$, the quadrupolar and magnetic broadenings due to charge inhomogeneity, which we attribute to CDW order in YBCO. We denote these with the superscript "cdw" to reflect their origin. However, it is important to note that, in the absence of identifiable features in the lineshape, NMR cannot determine the spatial scale of the charge modulation. In our case, the clear increase in linewidth at low temperature is attributed to static CDW correlations, since long-range variations in hole concentration (on nanometer to micrometer scales) are expected to be weak in YBCO and they are very unlikely to appear at low temperature.}

\textcolor{black}{The total quadrupole and magnetic contributions to the width, ${w}_{\text{quad}}$ and ${w}_{\text{magn}}$, can then be extracted from the width of the different lines as a function of temperature (Fig. \ref{width}a), using the following set of equations: }
\begin{widetext}
\begin{eqnarray}
w_{\text{quad}} &=& \sqrt{\left(w_{\text{quad}}^0\right)^2+\left(w_{\text{quad}}^{\rm cdw}\right)^2}
=\sqrt{ \frac{ w_{\text{HF2}}^2 + w_{\text{LF2}}^2 - \left( w_{\text{HF1}}^2 +w_{\text{LF1}}^2\right)}{6}}  \\
w_{\text{magn}} &=& \sqrt{ \left(w_{\text{magn}}^0\right)^2+\left(w_{\text{magn}}^{\rm cdw}\right)^2}
=\sqrt{ \frac{4(w_{\text{HF1}}^2+ w_{\text{LF1}}^2) - \left(w_{\text{HF2}}^2+w_{\text{LF2}}^2\right)}{6}}
\end{eqnarray}
\end{widetext}




\clearpage
\onecolumngrid

\section*{\large Supplementary Figures for "Robust Charge Density Wave Correlations in Optimally-Doped YBa$_2$Cu$_3$O$_{\rm{y}}$}



\begin{figure}[h!]
\includegraphics[width=7cm]{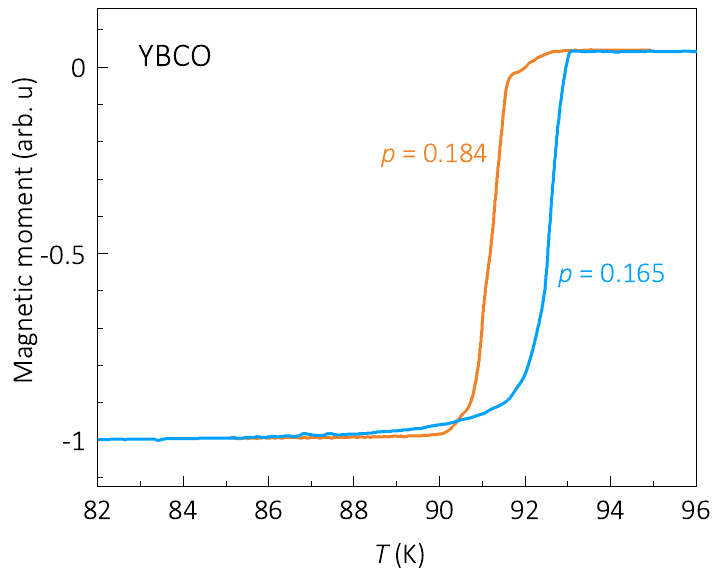}
  \caption{Temperature dependence of the magnetization across the superconducting transition $T_c$ for the two samples, as measured in a measured in a superconducting quantum interferometer device (SQUID).}
\label{squid}	
\vspace*{2cm}
\includegraphics[width=7cm]{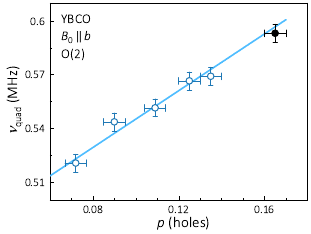}
  \caption{Doping dependence of the quadrupole frequency $\nu_{\rm quad}$ of O(2). The field is applied along the $b$ axis. Open circles are data from samples studied in refs.~\cite{Wu2015,Vinograd2019,Zhou2024}. Black dot: this work.}
\label{nuq-vs-p}	
\end{figure}

Fig.~\ref{squid}: Superconducting transition in magnetization data for the two samples.

Fig.~\ref{nuq-vs-p}: Doping dependence of the quadrupole frequency in YBa$_2$Cu$_3$O$_{\rm{y}}$.

\begin{figure}
\includegraphics[width=8cm]{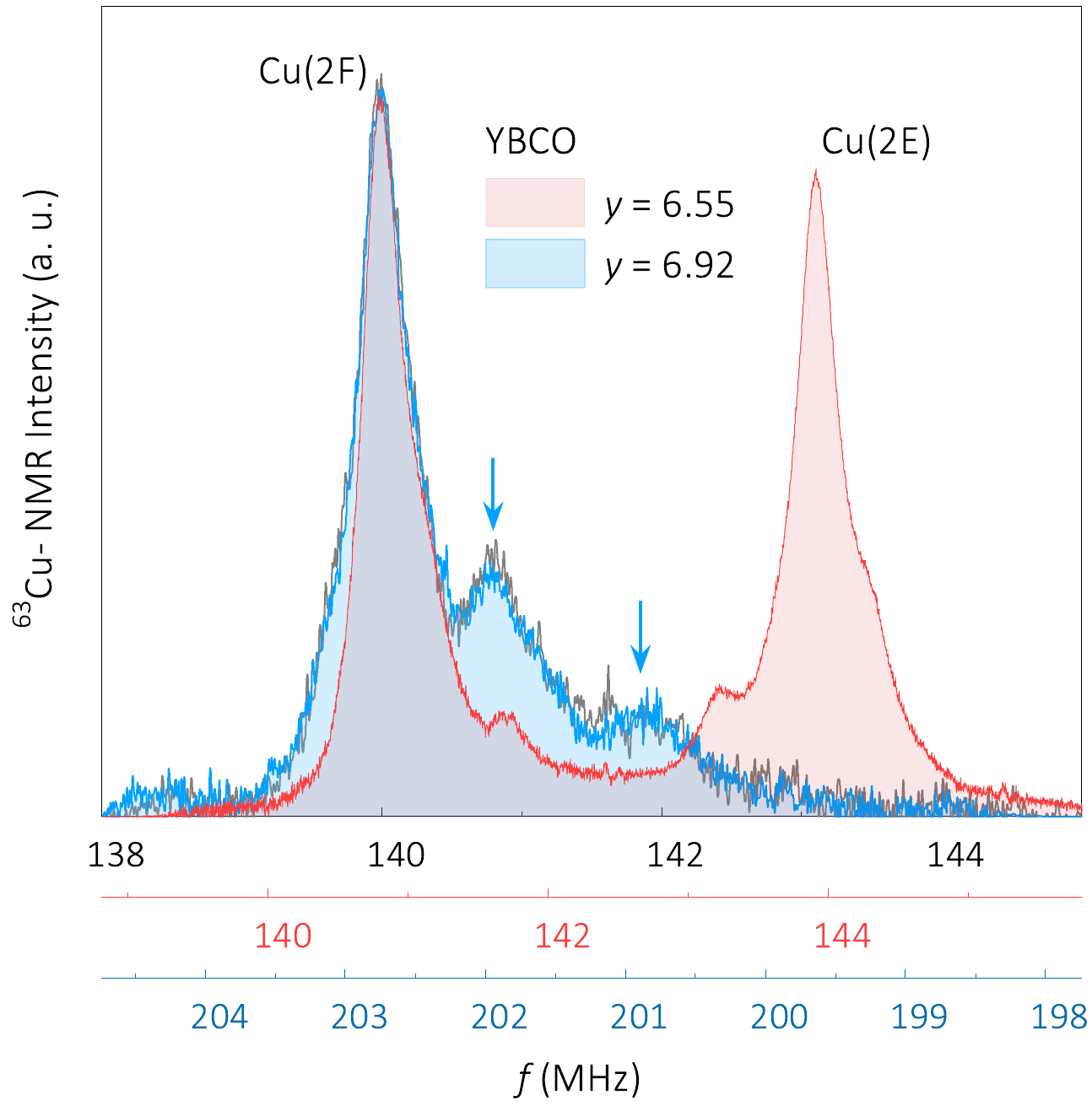}
  \caption{Planar $^{63}$Cu(2) quadrupole satellites in YBa$_2$Cu$_3$O$_{6.92}$ ($p=0.165$, this work,) and YBa$_2$Cu$_3$O$_{6.55}$ ($p=0.109$, from ref.~\cite{Wu2016}). For $p=0.165$, the low-frequency satellite (in gray) is shown on top of the (mirrored) high-frequency satellite (in blue). The absence of Cu(2E) line for $p=0.165$ is consistent with the ortho-I chain structure. The arrows show additional lines due to oxygen defects in the $p=0.165$ sample. Weaker additional lines, as well as shoulders in the main lines, are also seen in the $p=0.109$ sample and these are due to oxygen disorder (see discussion in ref.~\cite{Wu2016}).}
    \label{Cu-spectra}	
\vspace*{1cm}
\includegraphics[width=13cm]{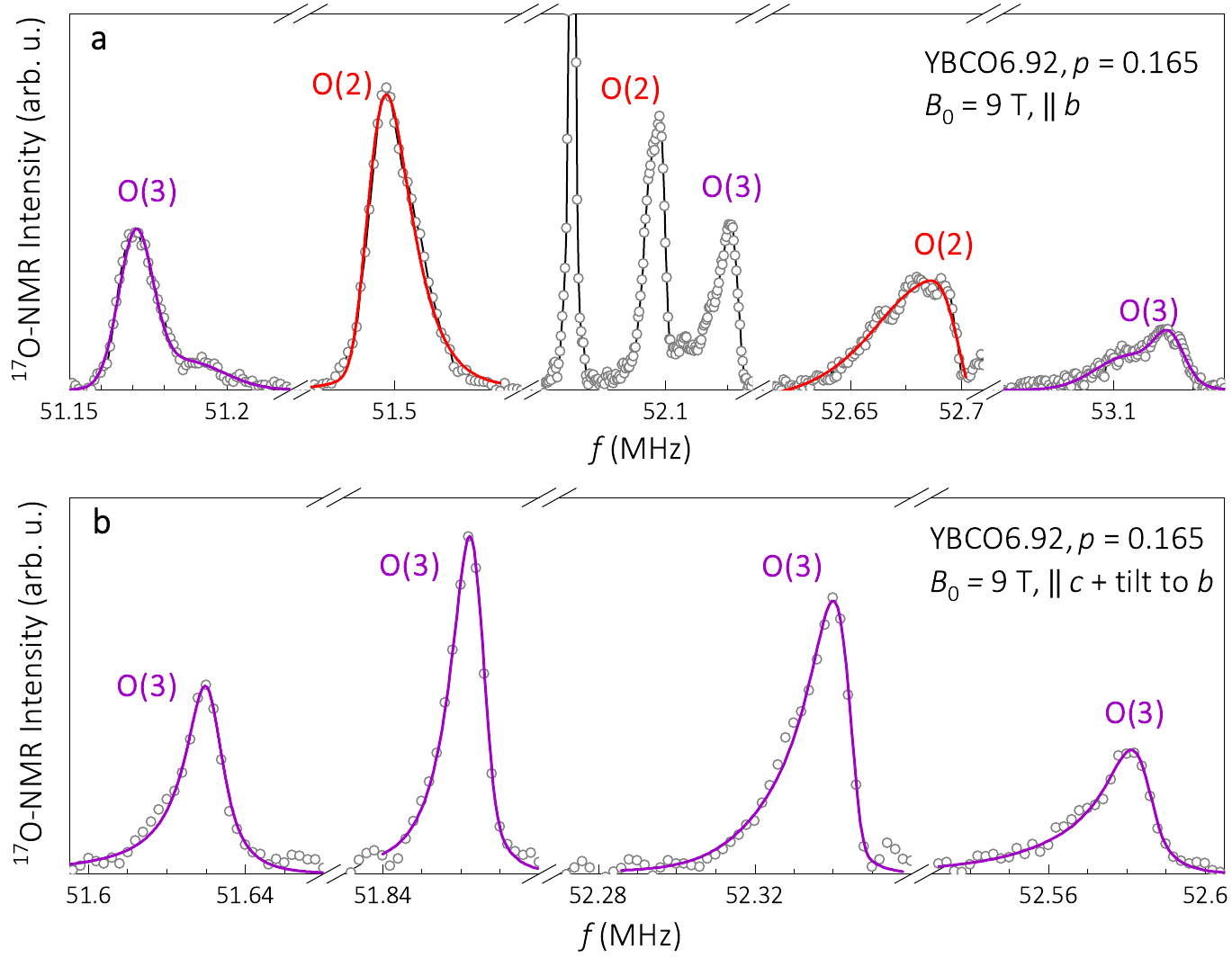}
  \caption{(a) $^{17}$O NMR spectrum for the $p = 0.165$ sample at $T = 170$~K and $B \parallel b$ axis (Open symbols). Only central lines and first quadrupole satellites are shown (second satellites are out-of-scale). The double peak structure for O(3) lines (not resolved on O(2) lines that are only asymmetric) is consistent with an effect due to chain-O vacancies that are closer to O(3) than to O(2). \textcolor{black}{To account for this double-peak structure, both O(2) and O(3) lines are fits to a pair of Gaussian functions}. (b) $^{17}$O(3) quadrupole satellites for $p = 0.165$ (open symbols) with the field tilted by 16$^\circ$ off the $c$ axis towards the $b$ axis. This \textcolor{black}{tilted} configuration offers the most favorable conditions for clean O(3) lineshapes \textcolor{black}{as the secondary peaks have shifted away (see their position in Fig.~\ref{spectra}a)}. Lines in panel (b) are fits to an asymmetric Pseudo-Voigt function.}
 \label{spectra-sup}	
\end{figure}

Fig.~\ref{Cu-spectra}: Comparison of $^{63}$Cu quadrupole satellites for y~=~6.55 and 6.92.

Fig.~\ref{spectra-sup}: $^{17}$O quadrupole satellites and fits to asymmetric peaks.

\begin{figure}
\vspace*{2cm}
\includegraphics[width=8.5cm]{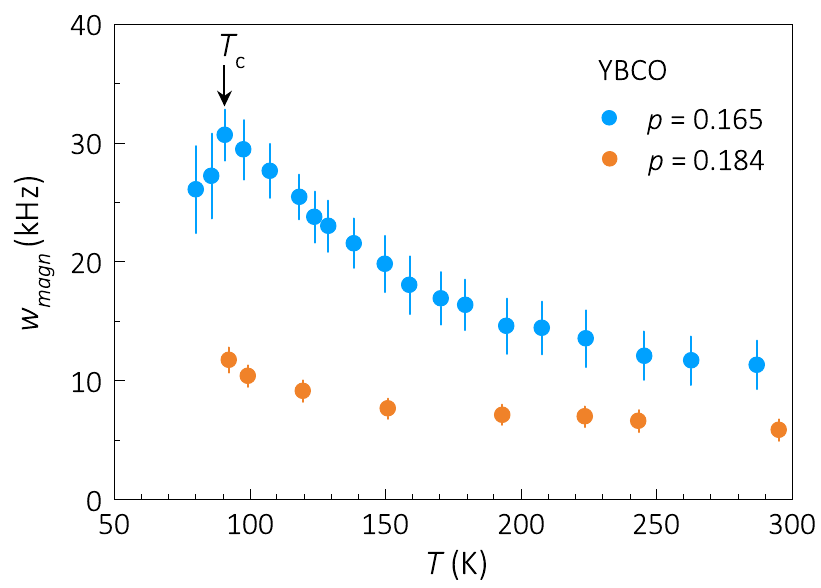}
  \caption{Temperature dependence of $w_{\rm magn}$ (see text) for $p=0.165$ and $p=0.184$ samples. $w_{\rm magn}(T)$ is contributed by both the CDW and the magnetic response to defects. This latter is a distribution of Knight shifts that widens upon cooling due staggered-magnetization puddles forming around non-magnetic defects~\cite{Julien2000,Ouazi2006,Alloul2009,Chen2009}}
\label{wmagn}	
\end{figure}

Fig.~\ref{wmagn}: Temperature dependence of the Knight shift contribution to the line width.

\begin{figure}
\includegraphics[width=8.5cm]{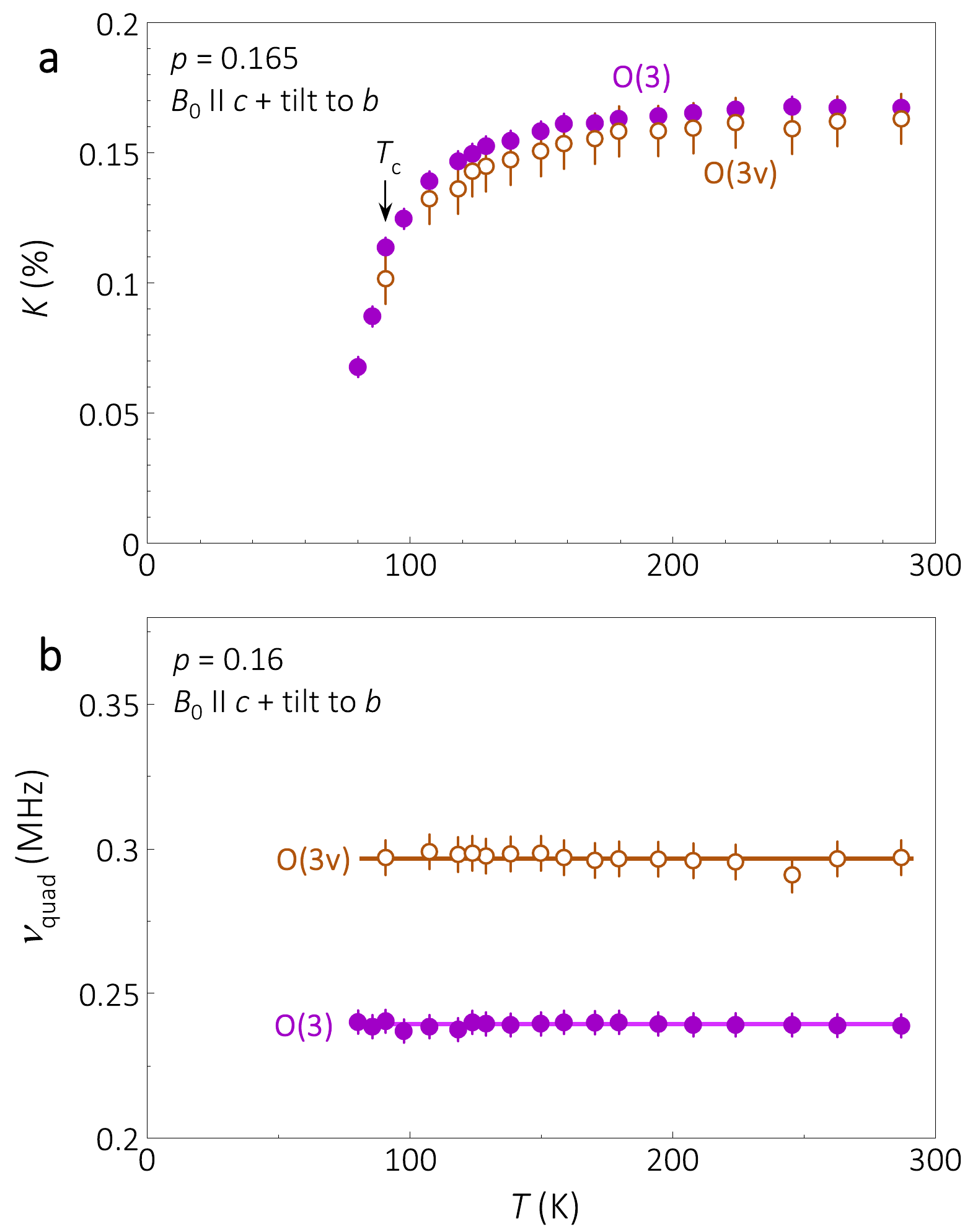}
  \caption{Temperature dependence of the Knight shift $K$ (a) and the quadrupole frequency $\nu_{\rm quad}$ (b) of O(3) and O(3v) sites in the $p = 0.165$ sample. The magnetic field is tilted by 16$^\circ$ off the $c$ axis towards the $b$ axis. }
        \label{nuq-vs-t}	
\end{figure}

Fig.~\ref{nuq-vs-t}: Temperature dependence of the Knight shift and the quadrupole frequency for O(3) and O(3v) sites.

\begin{figure}
\includegraphics[width=8.5cm]{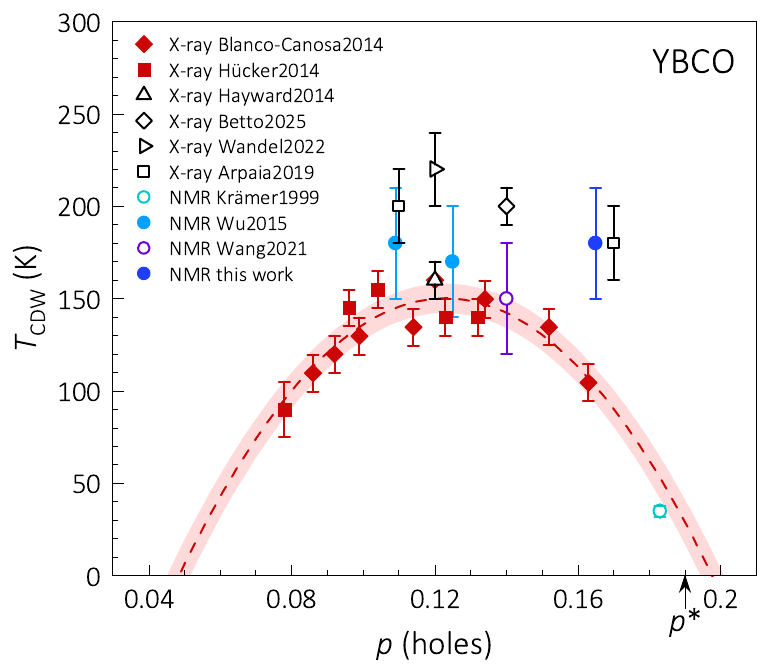}
  \caption{$T_{\rm CDW}$ from refs.~\cite{Huecker2014,Blanco-Canosa2014,Arpaia2019,Wu2015,Betto2025,Wandel2022,Kramer1999} vs. $p$. The dashed line is a fit of the data from refs.~\cite{Huecker2014,Blanco-Canosa2014} to a parabolic dependence and the shaded area depicts 2$\sigma$ standard deviation in the fit parameters. As discussed in the paper, there is no unambiguous definition of $T_{\rm CDW}$ for a single set of data and furthermore $T_{\rm CDW}$ may also depend on data analysis (peak intensity vs. integrated intensity, for example). Therefore, the quoted $T_{\rm CDW}$ values should not be taken at face value: they only represent the approximate temperature at which the CDW signal becomes identifiable in a given measurement with a given sensitivity. The purpose of this graph is to show that the CDW is unlikely to disappear around $p=0.16$.}
\label{parabola}	
\end{figure}

Fig.~\ref{parabola}: $T_{\rm CDW}~vs.~p$ values from the YBCO literature data.

\begin{figure}
\vspace*{2cm}
\includegraphics[width=17cm]{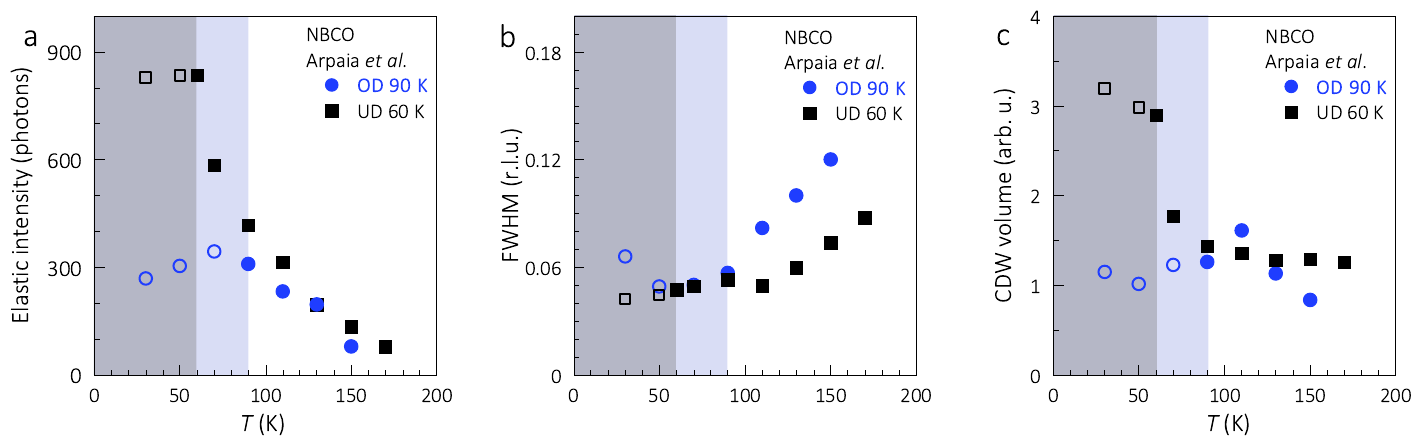}
  \caption{X-ray scattering data for underdoped (UD, $T_c=60$~K) and near optimally doped (OD, $T_c=90$~K) thin films of Nd$_{\rm 1+x}$Ba$_{\rm 2-x}$Cu$_3$O$_{\rm 7-\delta}$ (NBCO), adapted from Arpaia {\it et al.}~\cite{Arpaia2019}. (a) Intensity of the $(H,0)$ elastic peaks. (b) Width of the $(L,0)$ elastic peaks. (c) $q$-integrated peak intensity, i.e. peak intensity multiplied by the width in both $(H,0)$ and $(0,K)$ directions. Closed (open) symbols are used for data above (below) $T_c$. The shaded blue (grey) areas depict the superconducting states of the OD (UD) samples.}
        \label{arpaia}	
\end{figure}

Fig.~\ref{arpaia}: Comparison of X-ray scattering data of Arpaia {\it et al.}~for underdoped and overdoped NBCO.

\vspace{0.5cm}

\end{document}